\documentclass[8.5pt,twoside,twocolumn]{article}
\usepackage[super,sort&compress,comma]{natbib} 
\usepackage{mhchem}
\usepackage{times,mathptmx}
\usepackage{sectsty}
\usepackage{balance} 

\usepackage{graphicx} 
\usepackage{lastpage}
\usepackage[format=plain,justification=raggedright,singlelinecheck=false,font=small,labelfont=bf,labelsep=space]{caption} 
\usepackage{fancyhdr}
\begin{document}
\thispagestyle{plain}
\fancypagestyle{plain}{
\fancyhead[L]{\includegraphics[height=8pt]{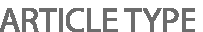}}
\fancyhead[C]{\hspace{-1cm}\includegraphics[height=20pt]{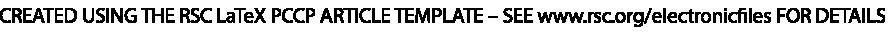}}
\fancyhead[R]{\includegraphics[height=10pt]{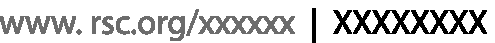}\vspace{-0.2cm}}
\renewcommand{\headrulewidth}{1pt}}
\renewcommand{\thefootnote}{\fnsymbol{footnote}}
\renewcommand\footnoterule{\vspace*{1pt}%
\hrule width 3.4in height 0.4pt \vspace*{5pt}} 
\setcounter{secnumdepth}{5}

\makeatletter 
\def\subsubsection{\@startsection{subsubsection}{3}{10pt}{-1.25ex plus -1ex minus -.1ex}{0ex plus 0ex}{\normalsize\bf}} 
\def\paragraph{\@startsection{paragraph}{4}{10pt}{-1.25ex plus -1ex minus -.1ex}{0ex plus 0ex}{\normalsize\textit}} 
\renewcommand\@biblabel[1]{#1}            
\renewcommand\@makefntext[1]%
{\noindent\makebox[0pt][r]{\@thefnmark\,}#1}
\makeatother 
\renewcommand{\figurename}{\small{Fig.}~}
\sectionfont{\large}
\subsectionfont{\normalsize}
\fancyfoot{}
\fancyfoot[LO,RE]{\vspace{-7pt}\includegraphics[height=9pt]{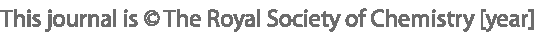}}
\fancyfoot[CO]{\vspace{-7.2pt}\hspace{12.2cm}\includegraphics{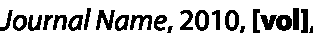}}
\fancyfoot[CE]{\vspace{-7.5pt}\hspace{-13.5cm}\includegraphics{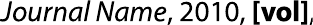}}
\fancyfoot[RO]{\footnotesize{\sffamily{1--\pageref{LastPage} ~\textbar  \hspace{2pt}\thepage}}}
\fancyfoot[LE]{\footnotesize{\sffamily{\thepage~\textbar\hspace{3.45cm} 1--\pageref{LastPage}}}}
\fancyhead{}
\renewcommand{\headrulewidth}{1pt} 
\renewcommand{\footrulewidth}{1pt}
\setlength{\arrayrulewidth}{1pt}
\setlength{\columnsep}{6.5mm}
\setlength\bibsep{1pt}

\twocolumn[
  \begin{@twocolumnfalse}
    \noindent\LARGE{\textbf{Dewetting and Spreading Transitions for Active
        Matter on Random Pinning Substrates}}
\vspace{0.6cm}

\noindent\large{\textbf{Cs. S\'{a}ndor,\textit{$^{1,2}$}, A. Lib\'{a}l,$^{\ast}$\textit{$^{1,2}$},
    C. Reichhardt,\textit{$^{2}$}, and
C. J. Olson Reichhardt\textit{$^{2}$} }}\vspace{0.5cm}

\noindent\textit{\small{\textbf{Received Xth XXXXXXXXXX 20XX, Accepted Xth XXXXXXXXX 20XX\newline
      First published on the web Xth XXXXXXXXXX 20XX}}}

\noindent \textbf{\small{DOI: 10.1039/b000000x}}
\vspace{0.6cm}

\noindent \normalsize{We show that sterically interacting self-propelled
disks in the presence of random pinning substrates 
exhibit transitions among a variety of different states.
In particular, from a phase separated cluster state, the disks
can spread out and homogeneously cover the substrate in what
can be viewed as an example of an active matter wetting transition.
We map the location of this transition as
a function of activity, disk density, and 
substrate strength, and we also identify other phases 
including a cluster state, coexistence between a cluster
and a labyrinth wetted phase, and a pinned liquid.
These phases can be identified
using the cluster size, which dips at the wetting-dewetting
transition, and the 
fraction of sixfold coordinated particles, which drops when
dewetting occurs.
}\vspace{0.5cm}
  \end{@twocolumnfalse}
]

\section{Introduction}

\footnotetext{\textit{$^{1}$~Faculty of Mathematics and Computer Science,
    Babe\c{s}-Bolyai University, Cluj, Romania 400084
    Fax: +40 264 591 906; Tel: + 40 264 405 300 /5240; E-mail: andras.libal@gmail.com}}
\footnotetext{\textit{$^2$~Theoretical Division and Center for Nonlinear Studies,
    Los Alamos National Laboratory, Los Alamos, New Mexico 87545, USA.}}

A wide class of systems exhibit pinning-induced 
order-disorder transitions in the presence of quenched disorder,
including vortices in type-II superconductors \cite{1,2},
two-dimensional (2D) electron crystals \cite{3,4}, charged colloids \cite{5,6,7}, 
soft matter systems with core-softened potentials \cite{8}, and 
hard disks \cite{9}.
When the ordered state is crystalline,
a transition to the disordered state as a function of increasing substrate strength
or decreasing particle density occurs through
the  proliferation of topological defects.
In addition to such equilibrium phases,
distinct phases can emerge under nonequilibrium conditions
in active matter or self-driven particle systems \cite{10,11},
including biological systems such as run-and-tumble bacteria \cite{12}
or artificial swimmers such as self-motile colloids \cite{11,13,14}.
One of the simplest
models of active matter is monodisperse sterically interacting 
disks undergoing  active Brownian motion or run and tumble dynamics.
These can transition
from a uniform liquid state to a clump or phase separated state
with increasing disk density,
increasing persistence length \cite{15,16,17,18,N,New1} or increasing
run length \cite{19,20,21}.
In the phase  separated regime,
which appears even in the absence of an attractive component in the
particle-particle interactions,
large clumps of densely packed disks are
separated by
a low density gas of active particles.
Monodisperse disks exhibit crystalline or polycrystalline ordering
within the high density regions inside the clumps \cite{16}.
A natural question to ask is how robust these cluster 
phases are in the presence of quenched disorder and
whether pinning-induced transitions
can occur as a function of increasing substrate strength.

Obstacle arrays, which have been considered in several studies
of swarming models \cite{22,23} and run and tumble disks \cite{21},
produce quite 
different effects from the collective behavior that can arise in pinning
arrays. The distinction between a pin and an obstacle is that it is possible
for particles to pass through a pinning site
either individually or collectively, while obstacles present an impenetrable
barrier.
The dynamics of many physically relevant active matter systems,
such as particles moving over rough substrates,
are better described in terms of an effective pinning landscape instead of
in terms of obstacle avoidance.
Studies of modified Vicsek models
in the presence of obstacles
showed that swarming was optimized
at a particular noise value \cite{22},
while in
other studies, increasing the disorder strength
caused a phase transition from a swarming to a non-swarming state \cite{24}.
In studies of self-propelled disks interacting 
with obstacle arrays, the mobility of the disks was
a non-monotonic function of the running length, since disks
with long running times spend more time trapped behind obstacles \cite{21}.

Here we consider
self-propelled disks interacting
with a substrate composed of randomly placed pinning sites.
A transition from a pin-free phase separated state to a homogeneous state
can be induced by increasing the substrate strength.    
This transition can be viewed as an active matter version 
of a wetting-dewetting or spreading transition \cite{25}, where the active
particles spread out to cover the surface when the pinning is strong.
We also show 
that a variety of different states can occur as function of disk density, substrate 
strength, and activity, including cluster phases, coexisting clustered and
wetted states,
a wetted percolating state, and a pinned liquid state.
These different states can be characterized by the size of the clusters and the 
amount of sixfold ordering of the disks.

\begin{figure}[t]
  \centering
\includegraphics[width=3.5in]{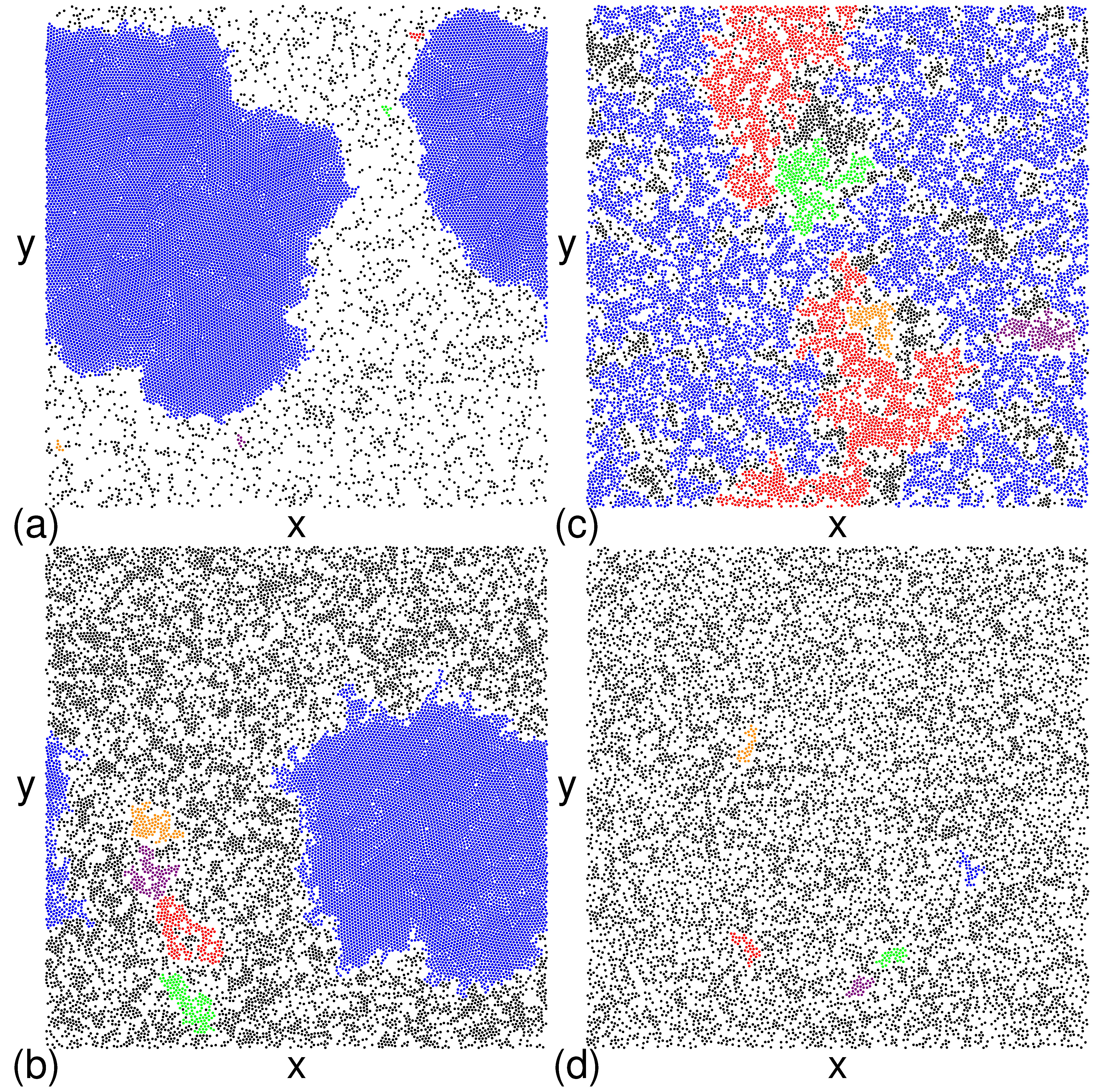}
\caption{ The disk positions (dots) for run and tumble disks interacting 
  with a random pinning substrate in samples with $l_r=300$ and
  $N_p/N_s=0.5$.
  Colors indicate
  the largest clusters in the system.
  (a) A dewetted state (Phase I) at $F_{p} = 1.0$ and $\phi=0.55$. (b)
  A partially wetted state (Phase II) at $F_{p} = 2.25$ and $\phi=0.55$.
  (c) At $F_{p} = 8.0$ and $\phi=0.55$, the disk density is uniform and
  the system forms a wetted state with disordered clusters (Phase III). 
(d) At $F_{p} = 8.0$ and $\phi = 0.349$, there is a pinned liquid state (Phase IV).
}
\label{fig:1}
\end{figure}

\section{Simulation}
We numerically simulate a 2D system of
$N_s=8000$ to 20,000 self-propelled disks using GPU based 
computing.
The disk radius is 
$R = 1.0$ and the system size is $L \times L$ with $L=300.0$,
giving a filling factor 
of $\phi =  \pi R^2/L^2 = 0.279$ to  $0.698$. The disks obey the following overdamped 
equation of motion:
\begin{equation}
\eta \frac{d{\bf r}_i}{dt} = {\bf F}_{\rm inter}^i + {\bf F}_m^i + {\bf F}_p^i ,
\end{equation}
where $\eta=1$ is the drag coefficient,
${\bf F}_{\rm inter}^i=\sum_{i\neq j}^{N_s}\Theta(d-2R)k(d-2R){\bf \hat d}$ is the 
repulsive disk-disk interaction force,
$d=|{\bf r}_i-{\bf r}_j|$, ${\bf \hat d}=({\bf r}_i-{\bf r}_j)/d$,
$k=20.0$ is the harmonic spring contact 
force, and $\Theta(x)$ is the Heaviside function.
The motor force
${\bf F}_m^i=F_m{\bf \hat m}_i$ with fixed $F_m=1.0$ acts
on each disk in a direction ${\bf \hat m}_i$
that changes randomly via a run and tumble protocol every
$t_r$
simulation time steps.
The time step used in the simulations is $dt=0.001$.
We characterize the activity of the disks by 
$\tilde l_{r}=F_mt_r dt$, which is the distance a disk would travel in a single
running time in the absence of disk-disk interactions or pinning, and take
$\tilde l_r$ to be uniformly distributed over the range
[$l_r,2l_r$].
${\bf F}_p^i$, the pinning force exerted by the substrate,
is modeled by an array of $N_p$
randomly 
placed circular parabolic traps with a finite radius of $R_p=0.5$,
${\bf F}_p^i=\sum_{k=1}^{N_p}F_p(r_p^{ik}/R_p)\Theta(r_p^{ik}-R_p){\bf \hat r}_p^{ik}$
where $F_{p}$ is the maximum 
pinning force exerted at the edge of the trap,
$r^{ik}=|{\bf r}_i-{\bf r}_k^{(p)}|$ is the distance 
from the center of disk $i$ to the center of pinning site $k$,
and ${\bf \hat r}_p^{ik}=({\bf r}_i-{\bf r}_k^{(p)})/r^{ik}$.
Since $R_p<R$, a given pinning site can trap no more than one disk at a time.

\begin{figure}
  \centering
\includegraphics[width=3.5in]{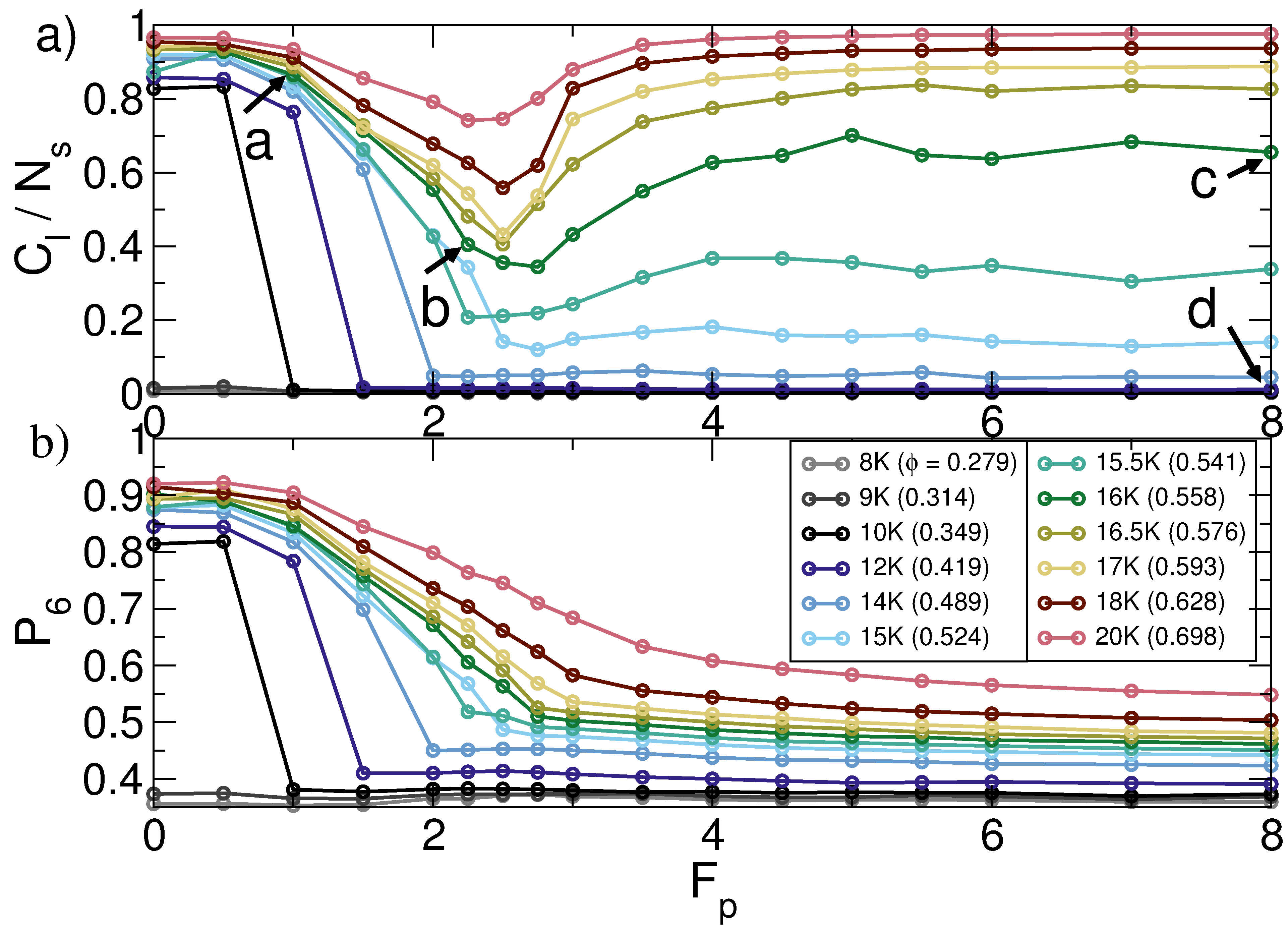}
\caption{ (a) The fractional size of the largest cluster $C_{l}/N_s$ vs $F_{p}$ at 
  $l_{r} = 300$ and $N_p=8000$ for
  $N_s=8000$ to 20,000 corresponding to $\phi=0.279$ to 0.698.
  The letters {\bf a}, {\bf b}, {\bf c}, and {\bf d}
  indicate the points at which the images in Fig.~\ref{fig:1}(a-d) were obtained.
  The dip near $F_p=2.5$ occurs at the transition from
  the dewetted phase I or the partially wetted phase II to the wetted phase III.
  (b) The corresponding fraction of sixfold coordinated particles $P_{6}$ vs $F_{p}$ shows
a drop as the system enters phase III.}
\label{fig:2}
\end{figure}

\section{Results}

In Fig.~\ref{fig:1} we show four representative images of the phases that appear
for active disks moving over a quenched pinning landscape in a sample
with $l_r=300$ and $N_p/N_s=0.5$.
The coloring highlights the largest individual clusters of disks, identified using
the algorithm of Luding and Herrmann 
\cite{26}. 
In the absence of a substrate, $F_p=0.0$, the disks form
a phase separated state
for these parameters.
For $F_p=1.0$ in Fig.~\ref{fig:1}(a),
a phase separated state containing a single high density cluster
is still present.  Disks in the surrounding low density gas state can be
temporarily pinned since $F_m=F_p$, but overall the morphology is
similar to that of the pin-free state.  We term this the active dewetted state, or Phase I.
At $F_{p} = 2.25$ in Fig.~\ref{fig:1}(b), a large
cluster is still present but numerous smaller clusters have nucleated due to
the trapping of gas phase disks by the pinning sites.
As a result, the large cluster is smaller than that shown in Fig.~\ref{fig:1}(a) while
the gas phase density is higher.
This partially wetted state,
called Phase II, can be viewed as a coexistence of the dewetted state,
consisting of the large cluster, and a wetted state, in which the particles
coat the entire substrate.
At $F_{p} = 8.0$ 
in Fig.~\ref{fig:1}(c), the single large cluster has vanished and the system adopts
a uniform labyrinth morphology which we refer to as the wetted state
or Phase III.
In general we 
observe a similar sequence of phases at lower disk densities but find that the
wetted state becomes less labyrinthine as the disks contact each other
less frequently,
as shown in Fig.~\ref{fig:1}(d) for $F_{p} = 8.0$ 
and $\phi = 0.349$ where the system forms a pinned liquid state called Phase IV.

\begin{figure}
  \centering
\includegraphics[width=3.5in]{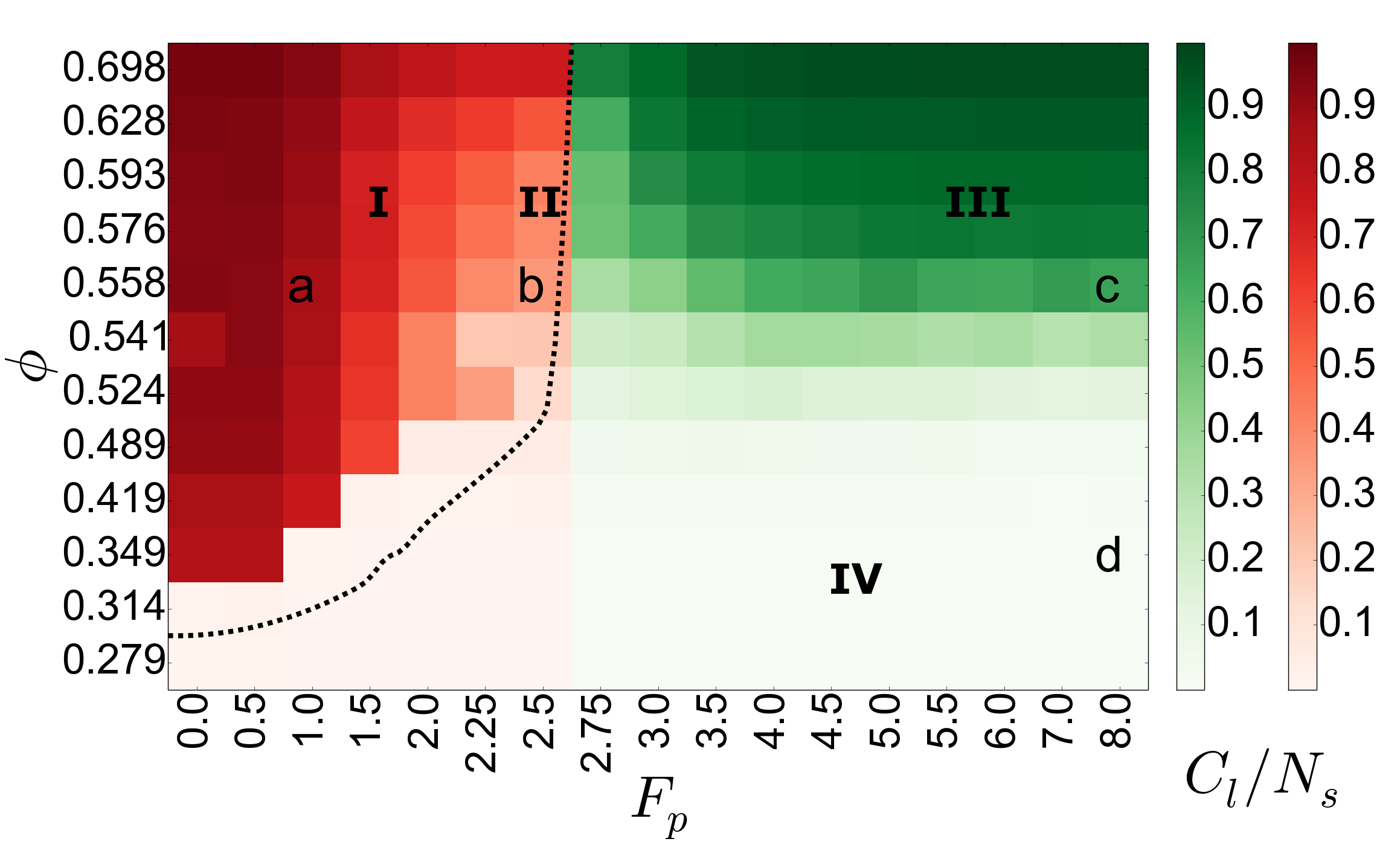}
\caption{ A heat map of $C_l/N_s$ showing the locations of the different phases
  as a function of $\phi$ vs $F_{p}$ for fixed $l_r=300$ and $N_p=8000$.
  Red areas for $F_p<2.75$ indicate the formation
  of large compact clumps, while in the green areas for $F_p \geq 2.75$, large
  branching clumps appear.
  I: dewetted phase; II: partially wetted phase (along dashed line);
  III: wetted phase; IV: pinned liquid.
  The letters {\bf a}, {\bf b}, {\bf c}, and {\bf d} indicate the values of
  $\phi$ and $F_p$ at which the images in Fig.~\ref{fig:1} were obtained.}
\label{fig:3}
\end{figure}

\begin{figure}
  \centering
\includegraphics[width=3.5in]{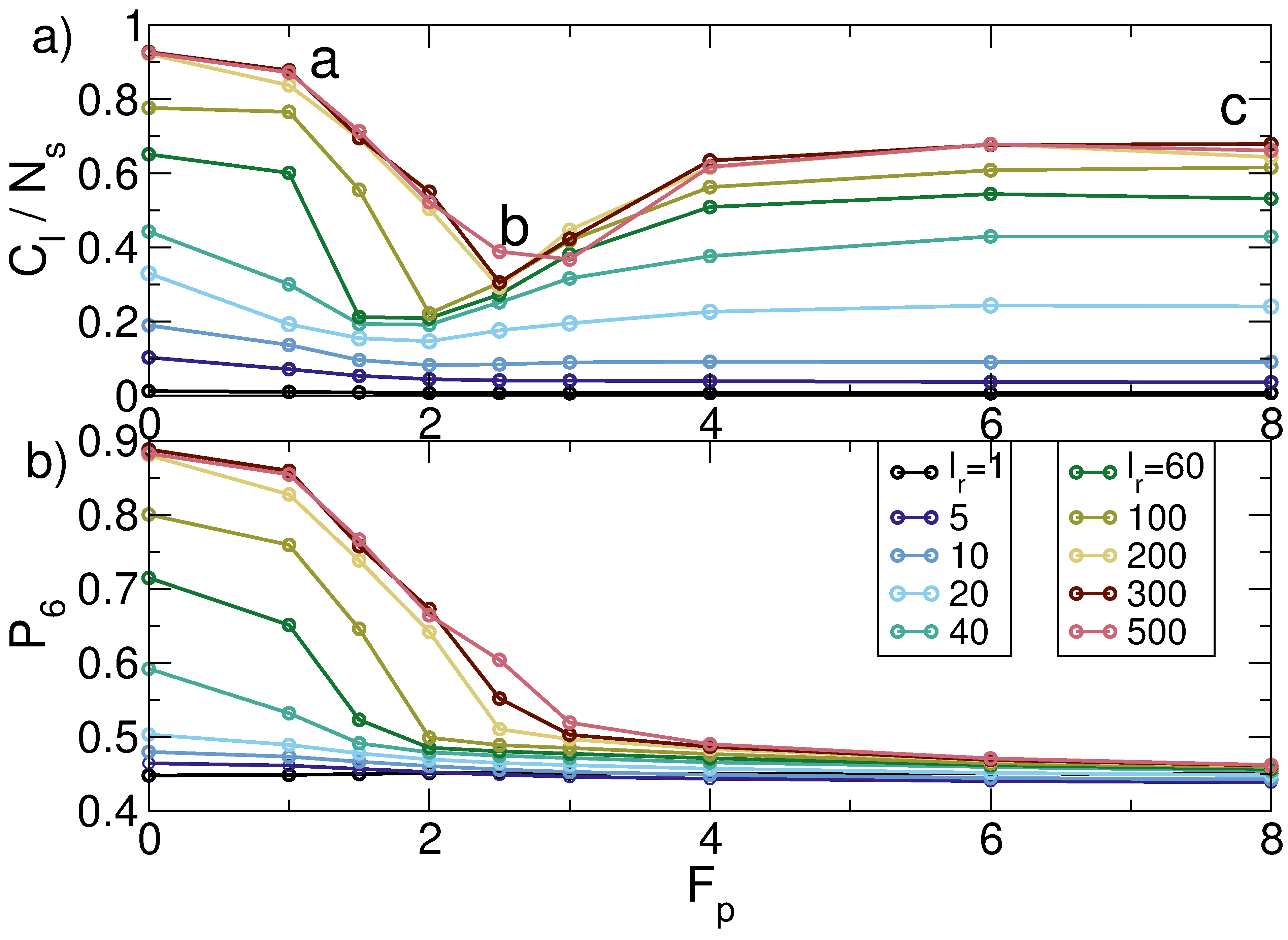}
\caption{ (a) The size of the largest cluster $C_{l}/N_s$ vs $F_{p}$ at fixed 
  $\phi = 0.55$ and $N_p/N_s=0.5$  for run lengths ranging from $l_r=1$ to $l_r=500$.
  (b) The corresponding fraction of sixfold coordinated particles $P_{6}$ vs
  $F_{p}$ shows a drop as the system enters phase III.}
\label{fig:4}
\end{figure}

In Fig.~\ref{fig:2}(a) we plot the fraction of particles in the largest cluster $C_{l}/N_s$ 
versus $F_{p}$ for the system in Fig.~\ref{fig:1} at a fixed run length of $l_r=300$
for varied $\phi$. Figure~\ref{fig:2}(b) shows the corresponding fraction $P_6$ of
sixfold coordinated disks obtained using the CGAL library\cite{30}.
For $\phi > 0.315$,
we find $C_l/N_s>0.8$ and $P_6>0.8$ at low $F_p$
since the system forms a
single large clump with strong sixfold ordering.
In the range $ 0.315 < \phi < 0.5$,
there is a pronounced drop in both $C_l/N_s$ and $P_6$ with increasing $F_p$ as the
system transitions from the
clump phase illustrated in Fig.~\ref{fig:1}(a) to
a pinned liquid phase of the type shown in Fig.~\ref{fig:1}(d).
For $\phi>0.5$,
just before $C_l/N_s$ reaches a minimum value at $F_p \approx 2.5$
the system enters a
partially wetted state similar to that shown in Fig.~\ref{fig:1}(b).
As $F_p$ increases further,
$C_{l}/N_s$ increases again but $P_6$ 
continues to decease, indicating that clusters with disordered structure have emerged,
as illustrated in Fig.~\ref{fig:1}(c) at $F_{p} = 8.0$
where the system forms a labyrinth state and the disk density becomes uniform.
The morphology of the large cluster is different in the two high $C_l/N_s$ regimes,
with a compact cluster forming in the dewetted state for $F_p < 2.5$, and a
much more porous, extended, and branching cluster appearing in the wetted state
for $F_p > 2.5$.
For $\phi < 0.5$ at high $F_p$, interconnections between small branching clusters
can no longer percolate across the sample, so there is no giant branching cluster and
$C_{l}/N_s$ remains low.
Overall, we find that for the dewetted cluster (I), $C_l/N_s$ and $P_6$ are both large
and the disk density is heterogeneous.
In the partially wetted phase (II), $C_l/N_s$ is low and $P_6$ has an intermediate
value while the disk density remains heterogenous.
The wetted labyrinth phase (III) has high $C_l/N_s$ and high $P_6$ along with a
homogeneous disk density.
Finally, in the pinned liquid phase (IV), $C_l/N_s$ and $P_6$ are both low and the disk density
is homogeneous.

In Fig.~\ref{fig:3}(a) we show a heat map \cite{31} based on $C_l/N_s$ values
as a function of $\phi$ versus $F_{p}$ indicating the locations of phases I through IV.
For $\phi < 0.35$, the system is too dilute to form clusters, so $C_l/N_s$
remains low at all values of $F_p$.

A dewetting-wetting transition from phase I to phase III occurs for $\phi \geq 0.35$,
with the dashed line indicating the sliver of partially wetted phase (II) that exists
close to this transition.
The transition from phase I to phase II is not sharply defined.
For the clump-forming densities $\phi \geq 0.35$, over the range
$0.0 < F_{p} < 2.75$ the radius $R_{\rm cl}$ of the compact clump decreases with
increasing $F_p$ while the density of the gaslike phase surrounding the clump
increases. 
A direct measurement of $R_{\rm cl}$
in the $\phi = 0.55$ sample gives $R_{\rm cl} \propto (F_c - F_{p})^\alpha$
with $\alpha = 1.0$ and $F_{c} = 2.75$.
In the dewetted phase I, there is a coexistence between a high density phase
inside the clumps in which the local density $\phi_h$ is close to the monodisperse
packing limit of $\phi_h \approx 0.9$, along with a low density phase with local density
$\phi_l \ll \phi_h$.  As $F_p$ increases, more disks become trapped by pinning sites, so that
the spatial extent of the dense phase decreases while
$\phi_h$ remains constant.  At the same time, $\phi_l$ increases until, at the transition
to the fully wetted phase III, $\phi_l=\phi$.

\begin{figure}
  \centering
\includegraphics[width=3.5in]{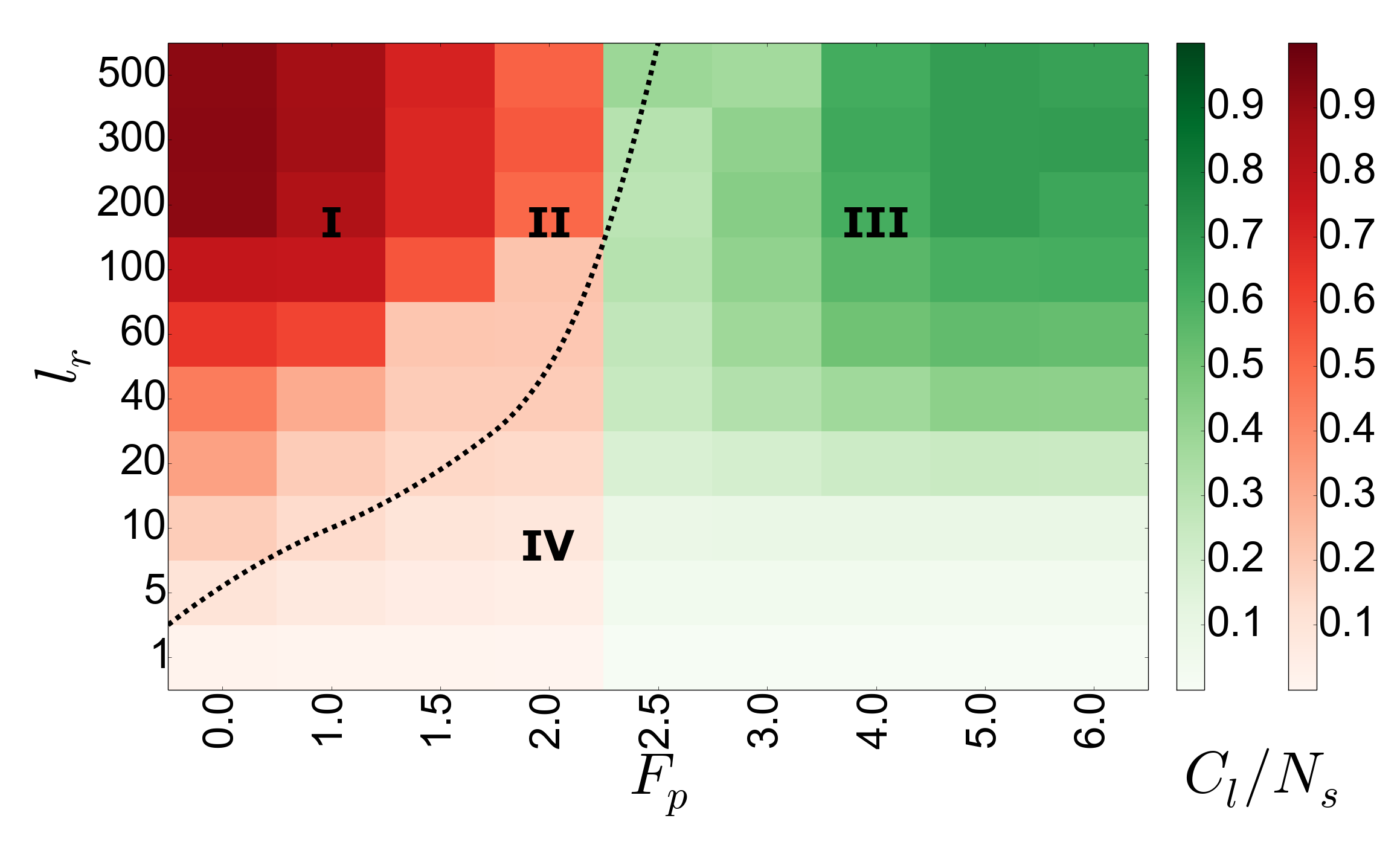}
\caption{A heat map of $C_{l}/N_s$ showing the locations of the
  different phases as a function of $l_{r}$ vs $F_{p}$ for fixed $\phi=0.55$
  and $N_p/N_s=0.5$.
  Red areas for $F_p<2.5$ indicate the formation
  of large compact clumps, while in the green areas for $F_p \geq 2.5$, large
  branching clumps appear.
  I: dewetted phase; II: partially wetted phase (along dashed line); III: wetted phase;
  IV: pinned liquid.
  }
\label{fig:5}
\end{figure}

We have also considered the effect of the run length by fixing
the disk density 
at $\phi=0.55$ and increasing $l_r$,
as shown in Fig.~\ref{fig:4}(a,b) where we plot $C_{l}/N_s$ and $P_{6}$ versus $F_{p}$.
For small $l_{r} < 20$,
$C_{l}/N_s$ and $P_{6}$ are both low and the system is in a pinned liquid state.
For large $l_{r} \geq 20$, a clump phase
appears for $F_{p} < 2.75$ and we observe a dip feature 
in $C_{l}/N_s$ and a drop in $P_6$ at the dewetting-wetting transition.
The overall behavior is very similar to that shown for varied $\phi$ and fixed $l_r$
in Fig.~\ref{fig:2}.
The heat map diagram of $C_l/N_s$ values in Fig.~\ref{fig:5}
as a function of $l_{r}$ versus $F_{p}$ illustrates
the locations of phases I through IV.

To test the effect of the pinning site density, we fix $l_r=300$, $\phi=0.55$, and
$F_p=2.0$ and increase the number of pinning sites $N_p$.
We find that at low pinning densities, a dewetted clump phase appears that transitions
to a wetted phase as $N_p$ increases.
One difference between sweeping $F_{p}$ and
sweeping $N_{p}$ is that at the highest pinning densities the 
percolating cluster phase disappears.  Since overlapping of pinning sites is
not allowed, trapped disks are not likely to come into contact with each other to form
a cluster, and at large $N_p$ almost every disk is trapped, so $C_l/N_s$ drops nearly
to zero.

\begin{figure}
  \centering
\includegraphics[width=3.5in]{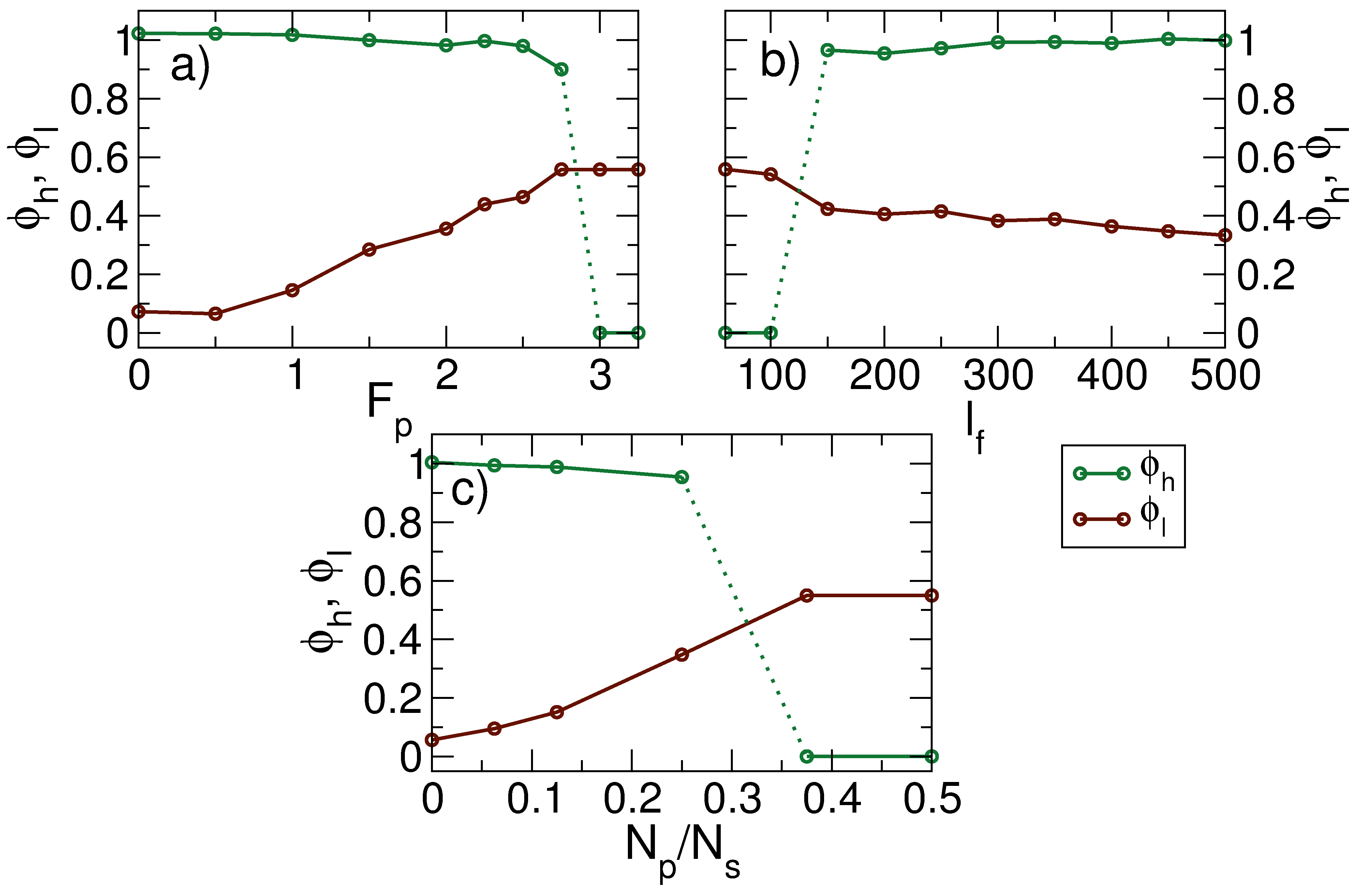}
\caption{ The local density $\phi_h$ inside the clusters (green) and
  $\phi_l$ in the gas phase (brown) at fixed $\phi=0.55$.
  (a) $\phi_h$ and $\phi_l$ vs $F_p$ for $l_r=300$ and $N_p/N_s=0.5$.
  (b) $\phi_h$ and $\phi_l$ vs $l_r$ for $F_p=2.0$ and $N_p/N_s=0.5$.
  c) $\phi_h$ and $\phi_l$ vs $N_p/N_s$ for $l_r=300$ and $F_p=5.0$.
Dashed lines indicate the point at which the large cluster disappears
from the system.}
\label{fig:6}
\end{figure}

In Fig.~\ref{fig:6} we plot the changes in the local density $\phi_h$ inside the clusters
and $\phi_l$ in the gas phase as a function of $F_p$, $l_r$, and $N_p/N_s$.
In each case, $\phi_h$ decreases slightly
from $\phi_h=0.9$ before suddenly dropping to $\phi_h=0$ when the cluster disappears
and the system reaches a uniformly wetted state.
At the same time, $\phi_l$ gradually increases as the transition to the
fully wetted cluster-free
state is approached.
The gentle decrease in $\phi_h$ in the cluster state occurs since
the effective pressure inside the cluster falls as the cluster shrinks.
The increase in $\phi_l$ is caused by a simple conservation of mass; as disks
leave the cluster they become part of the gas phase which fully wets the
substrate once $\phi_l=\phi$.

Our results could be tested using active mater systems in the presence 
of a rough substrate.  One method that can be used to create such a substrate 
is optical trapping, which allows the substrate strength to be tuned by
varying the light intensity.
There has already been some work examining the behavior of
run and tumble bacteria in optical trap arrays \cite{27,28}. 
Although we focus on run and tumble systems, our results should
be general to driven Brownian particle systems in which similar clustering 
transitions occur due to the density dependence of the particle motility \cite{29}. Since 
the disks in a cluster are less strongly coupled to the substrate 
than disks that are not part a cluster,  the onset of 
clustering may be a useful strategy
that could be exploited by living active matter to 
collectively escape from a disordered environment.

\section{Summary}

We have numerically examined run and tumble disks interacting with a 
random pinning substrate where we find that there can be active matter 
wetting-dewetting transitions as a function of pinning strength, 
disk density, and run length. In regimes where the pin-free system
forms a cluster state, we find that increasing the substrate strength
causes the size of the cluster to shrink gradually until the disk density
becomes homogeneous.  Here, the cluster phase can be viewed as a dewetted
state while the homogeneous phase is like a wetted state.
We show that the system exhibits different phases including
a clump state, a partially wetted state in which clumps coexist with a gas of pinned
disks,  a fully wetted labyrinth state, and a pinned liquid state.
Transitions between these states can be identified by measuring
the size of the largest cluster and 
the fraction of sixfold coordinated particles.
Our results indicate that 
pinning can induce transitions in the behavior of active matter systems
that are similar to the pinning-induced order-disorder transitions in equilibrium 
condensed matter systems.

\section{Acknowledgments}
This work was carried out under the auspices of the 
NNSA of the U.S. DoE at LANL under Contract No. DE-AC52-06NA25396. 
Cs. S\'{a}ndor and A. Lib\'{a}l
thank the Nvidia Corporation 
for their graphical card donation that was used in carrying out 
these simulations.

\footnotesize{

}


\begin{thebibliography}{99}

\bibitem{1}
  E.H. Brandt,
  {\it Rep. Prog. Phys.} 1995, {\bf 58}, 1465.

\bibitem{2}
  S.C. Ganguli, H. Singh, G. Saraswat, R. Ganguly, V. Bagwe, P. Shirage, A. Thamizhavel,
  and P. Raychaudhuri,
  {\it Sci. Rep.} 2015, {\bf 5}, 10613.

\bibitem{3}
M.-C. Cha and H.A. Fertig,
{\it Phys. Rev. Lett.} 1995, {\bf 74}, 4867.

\bibitem{4}
D. Carpentier and P. Le Doussal,
{\it Phys. Rev. Lett.} 1998, {\bf 81}, 1881.

\bibitem{5}
A. Pertsinidis and X.S. Ling,
{\it Phys. Rev. Lett.} 2008, {\bf 100}, 028303.

\bibitem{6}
C. Reichhardt and C.J. Olson,
{\it Phys. Rev. Lett.} 2002, {\bf 89}, 078301.

\bibitem{7}
S. Deutschl{\" a}nder, T. Horn, H. L{\" o}wen, G. Maret, and P. Keim,
{\it Phys. Rev. Lett.} 2013, {\bf 111}, 098301.

\bibitem{8}
E.N. Tsiok, D.E. Dudalov, Yu.D. Fomin, and V.N. Ryzhov,
{\it Phys. Rev. E } 2015, {\bf 92}, 032110.

\bibitem{9}
W. Qi and M. Dijkstra,
{\it Soft Matter} 2015 {\bf 11}, 2852.

\bibitem{10}
M.C. Marchetti, J.F. Joanny, S. Ramaswamy, T.B. Liverpool, J. Prost,
M. Rao, and R.A. Simha,
{\it Rev. Mod. Phys.} 2013, {\bf 85}, 1143.

\bibitem{11}
  C. Bechinger, R. Di Leonardo, H. L{\" o}wen, C. Reichhardt, G. Volpe,
  and G. Volpe,
  {\it Rev. Mod. Phys.}, in press (2016).

\bibitem{12}
H.C. Berg,
{\it Random Walks in Biology}
(Princeton University Press, Princeton, 1983).

\bibitem{13}
J.R. Howse, R.A.L. Jones, A.J. Ryan, T. Gough, R. Vafabakhsh, and
R. Golestanian,
{\it Phys. Rev. Lett.} 2007, {\bf 99}, 048102.

\bibitem{14}
G. Volpe, I. Buttinoni, D. Vogt, H.-J. K{\" u}mmerer, and C. Bechinger,
{\it Soft Matter} 2011, {\bf 7}, 8810.

\bibitem{15}
Y. Fily and M.C. Marchetti,
{\it Phys. Rev. Lett.} 2012, {\bf 108}, 235702.

\bibitem{16}
G.S. Redner, M.F. Hagan, and A. Baskaran,
{\it Phys. Rev. Lett.} 2013, {\bf 110}, 055701.

\bibitem{17}
J. Palacci, S. Sacanna, A.P. Steinberg, D.J. Pine, and P.M. Chaikin,
{\it Science} 2013, {\bf 339}, 936.

\bibitem{18}
I. Buttinoni, J. Bialk{' e}, F. K{\" u}mmel, H. L{\" o}wen, C. Bechinger, and T. Speck,
{\it Phys. Rev. Lett.} 2013, {\bf 110}, 238301.

\bibitem{N}
C. Tung, J. Harder, C. Valeriani, and  A. Cacciuto,
{\it Soft Matter} 2016, {\bf 12}, 555.

\bibitem{New1}
J. Bialk{\' e}, J.T. Siebert, H. L{\" o}wen, and T. Speck,
{\it Phys. Rev. Lett.} 2015, {\bf 115}, 098301.

\bibitem{19}
J. Tailleur and M.E. Cates,
{\it Phys. Rev. Lett.} 2008, {\bf 100}, 218103.

\bibitem{20}
A.G. Thompson, J. Tailleur, M.E. Cates, and R.A. Blythe,
{\it J. Stat. Mech.: Theor. Exp.} 2011, {\bf 2011}, P02029.

\bibitem{21}
C. Reichhardt and C. J. Olson Reichhardt,
{\it Phys. Rev. E} 2014, {\bf 90}, 012701.

\bibitem{22}
O. Chepizhko, E.G. Altmann, and F. Peruani,
{\it Phys. Rev. Lett.} 2013, {\bf 110}, 238101.

\bibitem{23}
O. Chepizhko and F. Peruani,
{\it Phys. Rev. Lett.} 2013, {\bf 111}, 160604.

\bibitem{24}
  D. Quint and  A. Gopinathan,
  {\it Phys. Biol.} 2015, {\bf 17}, 046008.

\bibitem{25}
D. Bonn, J. Eggers, J. Indekeu, J. Meunier, and E. Rolley,
{\it Rev. Mod. Phys.} 2009, {\bf 81}, 739.

\bibitem{26}
S. Luding and H.J. Herrmann,
{\it Chaos} 1999, {\bf 9}, 673.

\bibitem{30}
  M. Karavelas,
``2D Voronoi diagram adaptor'' in
{\it CGAL User and Reference Manual} (2016).

\bibitem{31}
  J.D. Hunter,
  {\it Comput. Sci. Eng.} 2007, {\bf 9}, 90.

\bibitem{27}
M Paoluzzi, R Di Leonardo, and L Angelani,
{\it J. Phys.: Condens. Matter} 2014, {\bf 26}, 375101.

\bibitem{28}
E. Pince, S.K.P. Velu, A. Callegari, P. Elahi, S. Gigan, G. Volpe, and G. Volpe,
{\it Nature Commun.} 2015, {\bf 7}, 10907.

\bibitem{29}
M.E. Cates and J. Tailleur,
{\it Europhys. Lett.} 2013, {\bf 101}, 20010.

\end{thebibliography}
\end{document}